\documentclass[sigconf]{acmart}
\usepackage[utf8]{inputenc} 
\usepackage[T1]{fontenc}    
\usepackage{url}            
\usepackage{booktabs}       
\usepackage{amsfonts}       
\usepackage{nicefrac}       
\usepackage{microtype}      
\usepackage[page,title]{appendix}
\usepackage{comment}
\usepackage{enumitem}
\usepackage{threeparttable}
\usepackage{bm}
\usepackage{dsfont}
\usepackage{amsmath}
\usepackage{amsthm}
\usepackage{graphicx}
\usepackage{subfigure}
\usepackage{capt-of}
\usepackage{hyperref}
\usepackage{comment}

\usepackage{wrapfig}      
\usepackage{diagbox}
\usepackage{multirow}
\usepackage{makecell}

\usepackage{pifont}
\usepackage[ruled]{algorithm2e}

\AtBeginDocument{%
  \providecommand\BibTeX{{%
    \normalfont B\kern-0.5em{\scshape i\kern-0.25em b}\kern-0.8em\TeX}}}

\copyrightyear{2022} 
\acmYear{2022} 
\setcopyright{acmlicensed}
\acmDOI{10.1145/3447548.3467413}

\begin{document}

\author{Quanyu Dai$^{1}$, Zhenhua Dong$^{1}$, Xu Chen$^{2,3,*}$}\thanks{$*$ Corresponding author.}
\affiliation{
\institution{$^1$Huawei Noah's Ark Lab, $^2$Beijing Key Laboratory of Big Data Management and Analysis Methods, $^3$Gaoling School of Artificial Intelligence Renmin University of China}
\city{}
\country{}
} 

\title{Debiased Recommendation with Neural Stratification}

\begin{abstract}
Debiased recommender models have recently attracted increasing attention from the academic and industry communities. 
Existing models are mostly based on the technique of inverse propensity score (IPS).
However, in the recommendation domain, IPS can be hard to estimate given the sparse and noisy nature of the observed user-item exposure data.
To alleviate this problem, in this paper, we assume that the user preference can be dominated by a small amount of latent factors, and propose to cluster the users for computing more accurate IPS via increasing the exposure densities. 
Basically, such method is similar with the spirit of stratification models in applied statistics.
However, unlike previous heuristic stratification strategy, we learn the cluster criterion by presenting the users with low ranking embeddings, which are future shared with the user representations in the recommender model.
At last, we find that our model has strong connections with the previous two types of debiased recommender models.
We conduct extensive experiments based on real-world datasets to demonstrate the effectiveness of the proposed method.
\end{abstract}



\maketitle
\sloppy

\section{Introduction}
Recommender systems have been widely deployed in a large amount of applications, ranging from the news websites, music apps to the video sharing platforms and e-commerce websites. 
Traditional recommender models are mostly trained based on the observational data, which can be skewed due to the exposure~\cite{ExpoMF,CIKM_YuanHYZCDL19} or selection~\cite{IPS,DRJL} bias.
For correcting such biases, the inverse propensity score (IPS) is a mainstream technique~\cite{IPS,DRJL,CIKM_YuanHYZCDL19,Multi_IPW}, where the basic idea is to adjust the sample weights according to the observational probability of the user-item pairs.
While this method has achieved remarkable successes, accurately estimating the observational probability in the recommendation domain is not easy, since the user-item interactions are highly sparse and noisy.

\begin{figure}[t]
\centering
\setlength{\fboxrule}{0.pt}
\setlength{\fboxsep}{0.pt}
\fbox{
\includegraphics[width=.95\linewidth]{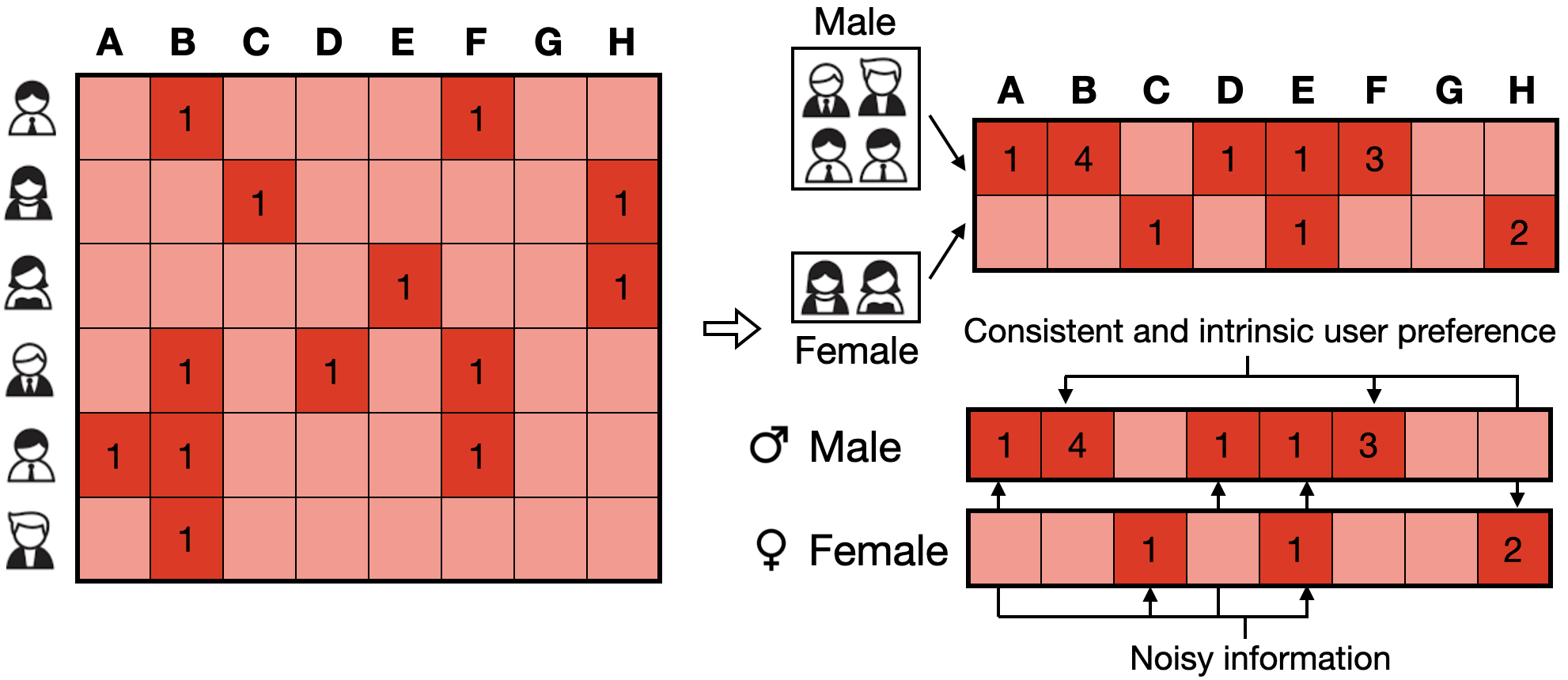}
}
\caption{Illustration of the advantages of cluster-level IPS.}
\label{intro}
\end{figure}

Intuitively, in a real-world recommender system, similar users may interact with the items with similar probabilities, for example, the science-fiction fans may all like ``The Matrix'', but do not prefer ``Forrest Gump'', which means, for all these users, ``The Matrix'' may have consistently higher observational probability.
This intuition inspires us to firstly cluster similar users, and then leverage the cluster-level IPS to debias the recommender model.
We argue that the advantages of such method lie in two aspects:
to begin with, the cluster-item observational probability can be derived based on more denser data.
As exampled in Figure~\ref{intro}, in the left user-item interaction matrix, each user interacts with at most three items.
By clustering the users according to their genders, the gender-item matrix are much more denser, which facilitates more accurate IPS estimation.
And then, by aggregating the interactions of similar users in the same cluster, the consistent and intrinsic user preferences can be highlighted, and the random noisy information can be simultaneously weakened.
Also see the example in Figure~\ref{intro}, the male users all interact with item B, which means item B can reflect the basic preference of this user group.
As expected, by aggregating the interactions, the weighting of item B is enhanced (see the right bottom  sub-figure).
On the contrary, since the noisy user behaviors are usually random, they do not concentrate on fixed items and cannot be highlighted by the aggregating operation.

Actually, the above clustering idea share some similarities with the stratification\footnote{In this paper, the terms stratification and clustering are used interchangeably.} methods in traditional causal inference literature.
Conventional stratification strategies cluster the confounders (\emph{i.e.}, users) according to some pre-defined features.
However, without enough priors, it is hard to determine which features are optimal.
To alleviate this problem, we leverage deep clustering to automatically learn the cluster criterion by projecting the raw user features into a latent space.
For connecting the clustering process with the recommendation task, we equalize the above user latent representations with the user embeddings in the recommender model.

Notably, there are some previous work~\cite{yang2022debiased} on debiased recommendation with representation learning.
While they also represent users with latent embeddings, they do not cluster the users, which significantly differs from our idea.
Actually, by specifying different cluster numbers, we find that our model has strong connections with the previous IPS and representation learning based debiased recommender models.

The main contributions of this paper can be summarized as follows: 
(1) We propose to perform user stratification with deep clustering for accurately estimating the propensity score, and deriving more effective unbiased recommender models. 
(2) To achieve the above idea, we design an algorithm by connecting the neural clustering and recommender models via sharing the user embeddings.
(3) We analyze the connections between our model and the previous IPS and representation learning based debiased models.
(4) We conduct extensive experiments based on real-world datasets to thoroughly evaluate the proposed method.

\section{Preliminary}
In a typical recommendation problem, we usually have a user set $\mathcal{U}=\{u_1,u_2,...,u_N\}$, an item set $\mathcal{V}=\{v_1,v_2,...,v_M\}$ and a user-item interaction set $\mathcal{O} = \{(u,v,o_{uv})\}$, where each element $(u,v,o_{uv})$ means there is an interaction between user $u$ and item $v$, and the feedback is $o_{uv}$.
If user $u$ has positive feedback on item $v$, then $o_{uv}=1$, otherwise $o_{uv}=0$.
Suppose the recommender model is $g$, then the user-item preference is predicted as:
\begin{equation}\label{g}
    \hat{o}_{ui} = g(\boldsymbol{s}_u, \boldsymbol{s}_i;\boldsymbol{\theta}_{g}),
\end{equation}
where $\boldsymbol{s}_u$ and $\boldsymbol{s}_i$ are the representations of the user $u$ and item $v$.
In practice, they can be derived based on the ID or profile information.
$\boldsymbol{\theta}_{g}$ are the model parameters.
For learning $g$, traditional models usually optimize the following target:
\begin{equation}
   L_{b} = \frac{1}{|\mathcal{O}|}\sum_{(u,v): (u,v,o_{uv})\in \mathcal{O}}e_{uv},
\end{equation}
where $e_{uv}$ can be implemented with either RMSE or binary cross entropy.
For RMSE, $e_{uv} = (o_{uv} - \hat{o}_{ui})^2$.
For binary cross entropy, $e_{uv} = o_{uv}\log{\hat{o}_{ui}} + (1-o_{uv})\log{(1-\hat{o}_{ui})}$.

If we consider the ideal learning objective of a recommender model, we should optimize the user feedback on each item~\cite{IPS}, that is:
\begin{equation}
   L_{\text{ideal}} = \frac{1}{|\mathcal{U}||\mathcal{V}|}\sum_{u\in \mathcal{U}} \sum_{v\in \mathcal{V}}e_{uv}.
\end{equation}
As pointed out by the previous work~\cite{IPS,chen2020bias}, $L_{b}$ is biased from $L_{\text{ideal}}$ since the observational dataset $\mathcal{O}$ can be influenced by the exposure or selection bias~\cite{chen2020bias}.
Unfortunately, directly optimizing $L_{\text{ideal}}$ is intractable, since we cannot obtain the user feedback on all items.
To build feasible unbiased recommender models, an important technique is the inverse propensity score, based on which the traditional optimization target $L_{b}$ is improved to:
\begin{equation}
    \mathcal{L}_{\text{ips}} = \frac{1}{|\mathcal{U}||\mathcal{V}|} \sum_{(u,v,o_{uv})\in\mathcal{O}}\frac{e_{uv}}{\hat{p}_{uv}},
\end{equation}
where $\hat{p}_{uv}$ is the probability of observing the interaction between user $u$ and item $v$.
Obviously, this method is highly depend on the accurate estimation of $\hat{p}_{uv}$.
However, in the recommendation domain, this is not easy, since the user-item interactions can be quite sparse and noisy.
In this paper, we propose to cluster the users to densify the data and resist the noisy information, which facilitates more accurate IPS estimation and more effective debiasing models (we call our model as C-IPS).

\section{The C-IPS Model}
In general, our model includes three phases:
(1) \textit{User clustering}, where the raw user features are projected into a latent space, based on which similar users are grouped into the same cluster.
(2) \textit{Cluster-level IPS estimation}, where we regard the user interactions in the same cluster as the cluster interactions, and compute the cluster-level IPS. 
(3) \textit{Debiased model learning}, where we optimize the recommender model by re-weighting the training samples with the above cluster-level IPS.
In phase (1) and (3), we share the user representations to make the clustering process more targeted.
In the following, we detail each of the above phases.

\subsection{User Clustering}
As mentioned above, without enough prior knowledge, it is hard to determine the optimal clustering criterion (\emph{i.e.}, computing user similarities based on which features).
Thus, we borrow the idea of deep clustering, which automatically learns to cluster the users.

Formally, let $\bm{x}_u$ be the raw features\footnote{In our model, the user raw features are represented by a $|\mathcal{V}|$-dimensional 0-1 vector, which indicates the items interacted by the user.} of user $u$, and we project them into an embedding by $\bm{h}_u = \phi(\bm{x}_u)$.
The key idea of deep clustering is to simultaneously learn the projection function $\phi$ and the user cluster assignments~\cite{DEC}.
More specifically, the probability of assigning a user $u$ to a cluster $k$ is computed as follows:  
\begin{equation}\label{aij}
    a_{uk} = \frac{(1+||\boldsymbol{h}_u - \boldsymbol{\mu}_k||)^{-1}}{\sum_{i=1}^K(1+||\boldsymbol{h}_u - \boldsymbol{\mu}_i||)^{-1}},
\end{equation}
where $K$ is the total number of clusters, and $\boldsymbol{\mu}_k$ is the learnable center of the $k$th cluster.
This equation holds the premise that if the user embedding $\bm{h}_u$ is more loser to $\boldsymbol{\mu}_k$, then the user is more likely to be grouped into the $k$th cluster.

To learn $\phi$ and the cluster centers $\{\boldsymbol{\mu}_k\}$, the commonly used KL divergence is leveraged to minimize the distance between the user cluster assignments and a target distribution $\bm{t}_{u}$, that is:
\begin{equation}\label{loss_dec}
    \mathcal{L}_{cluster} = \sum_{u\in \mathcal{U}} KL(\bm{t}_{u}||\bm{a}_u) = \sum_{u\in \mathcal{U}} \sum_{k=1}^K t_{uk}\log\frac{t_{uk}}{a_{uk}},
\end{equation}
where the target distribution $\bm{t}_{u}$ is specified as follows~\cite{DEC}:
\begin{equation}\label{tij}
    t_{uk} = \frac{a_{uk}^2/s_k}{\sum_{i=1}^K a_{ui}^2/s_i},
\end{equation}
where $s_i = \sum_i a_{ui}$, and such design is to encourage the clustering model to learn from its own high confidence predictions as in self-training~\cite{Co_Training,DEC}.
We summarize the training process in Algorithm~\ref{dec-alg}.

\begin{algorithm}[t]
\caption{{Learning algorithm for user clustering}}
\label{dec-alg} 
Initialize the parameters of $\phi$ and the cluster centers $\{\boldsymbol{\mu}_k\}$.\\
\For{epoch number in [1,~T]}{ 
    Compute the user cluster assignments via Eq.~(\ref{aij}).\\
    Update the target distribution via Eq.~(\ref{tij}).\\
    \For{u in $\mathcal{U}$}{
        Compute the cluster assignments of $u$ via Eq.~(\ref{aij}).\\
        Update $\phi$ and $\{\boldsymbol{\mu}_k\}$ via Eq.~(\ref{loss_dec}).\\
    }
}
\end{algorithm}

\subsection{Cluster-level IPS estimation}
Based on the above clustering results, in this phase, we compute the propensity scores.
Intuitively, similar users should interact the same item with similar probabilities.
Thus, we compute a unified cluster-level propensity score for all the users in the same cluster, that is:
\begin{equation}\label{clust_ps}
    \hat{p}_{uv} = \frac{ \sum_{u' \in \mathcal{U}} \bm{1}(c_{u'} = c_{u}) \bm{1}((u',v,o_{u'v})\in \mathcal{O}) }{ \max_{v'\in\mathcal{V}} \sum_{u' \in \mathcal{U}} \bm{1}(c_{u'} = c_{u}) \bm{1}((u',v',o_{u'v'}) \in \mathcal{O})},
\end{equation}
where $c_u$ is the cluster index of user $u$, that is, $c_u= \arg\max_j a_{uj}$. ``$\bm{1}(\text{condition})$'' is the indicator function, which is 1 if the condition holds, and 0 otherwise.
This equation actually compute the ratio between the interaction frequency of item $v$ in group $c_u$ and the maximum frequency across different clusters.
It should be noted that the propensity score can also be computed by more advanced method based on the cluster-item exposure matrix, and we left them as future work.

\subsection{Debiased model learning}
Once we have derived the propensity score, the recommender model can be optimized based on the following objective:
\begin{equation}\label{loss_hard}
    \mathcal{L}_{rec} =\!\!\!\sum_{(u,v, o_{uv})\in\mathcal{O}}\!\!\!\!\!\!\! \frac{o_{uv}\log{(\bm{s}_u)^T\bm{s}_v} + (1-o_{uv})\log{(1-(\bm{s}_u)^T\bm{s}_v)} }{\hat{p}_{uv}},
\end{equation}
where we learn the recommender model based on binary cross entropy.
$\bm{s}_u$ and $\bm{s}_v$ are the embeddings of user $u$ and item $v$. 
For connecting the clustering process with the recommendation task, we explicitly let $\bm{s}_u = \bm{h}_u$.

The complete learning process of our model is summarized in Algorithm~\ref{alg-all}.
In each iteration, we firstly learn the user clusters by Algorithm~\ref{dec-alg}, and then the propensity scores are computed based on the clustering results, based on which the final recommender model is optimized by re-weighting different samples.
We have also tested the joint learning strategy, but the performance is suboptimal.

\begin{algorithm}[t]
\caption{Learning process of ClusterIPS} 
\label{alg-all} 
Initialize the parameters of $\phi$, the cluster centers $\{\boldsymbol{\mu}_k\}$, the item embeddings $\{\bm{s}_v\}$.\\
\For{iteration number in [1,~M]}{
    Learn the user clusters based on Algorithm~\ref{dec-alg}.\\
    Compute the propensity scores based on equation~(\ref{clust_ps}).\\
    Learn the recommender model based on equation~(\ref{loss_hard}).
}
\end{algorithm}

\subsection{Discussion}
Our model borrows the stratification idea in traditional causal inference literature, which is an effective method for computing the causal effect.
However, we do not directly migrate the previous work, but try to adapt it to the recommendation domain, where we propose to automatically learn the stratification features.
If one closely inspect our model, she may find that it can somehow connect the previous two types of debiased recommender models.
In specific, if we set the cluster number as the user number (\emph{i.e.}, $K = |\mathcal{U}|$), then our model is reduced to the original user-level IPS methods.
If we set the cluster number as one, that is all the users are forced to have similar embedding distributions, then our model has similar effect\footnote{In this case, we realized that our model cannot be rigorously reduced to the balancing based models, but we would like to highlight that there are some connections.} as the representation learning based debiased recommender models~\cite{CBR}.

\begin{table*}[t]
\caption{Performance comparison between our model and the baslines. The best results are highlighted in bold.}
\renewcommand\arraystretch{1.2}
\center
\begin{tabular}{c|cccccccccc}
\hline\hline
   & Models                   & DCG@1           & Recall@1        & MAP@1           & DCG@3           & Recall@3        & MAP@3           & DCG@5           & Recall@5        & MAP@5           \\
\hline
\multirow{6}{*}{All} & MF                      & 0.0874          & 0.0989          & 0.0865          & 0.2321          & 0.2978          & 0.1816          & 0.3098          & 0.4913          & 0.2460          \\
                     & WMF                     & 0.1196          & 0.1195          & 0.1187          & 0.2609          & 0.3176          & 0.2147          & 0.3354          & 0.5043          & 0.2793          \\
                     & ExpoMF                  & 0.1481          & 0.1378          & 0.1472          & 0.2963          & 0.3392          & 0.2533          & 0.3739          & 0.5330          & 0.3204          \\
                     & Rel-MF                  & 0.1816          & 0.1623          & 0.1807          & 0.3748          & 0.3927          & 0.3241          & 0.4578          & 0.5890          & 0.4057          \\
                     \cline{2-11}
                     & C-IPS          & \textbf{0.2073}          & \textbf{0.1814}          & \textbf{0.2065}           & \textbf{0.4005}           & \textbf{0.4069}           & \textbf{0.3554}           & \textbf{0.4778}           & \textbf{0.5938}           & \textbf{0.4337}           \\
                    
                     \hline\hline
\multirow{6}{*}{Non-Popular} & MF                      & 0.0817          & 0.1044          & 0.0807         & 0.2197          & 0.3161          & 0.1708          & 0.2932         & 0.5215         & 0.2316          \\
                      & WMF                     & 0.0952          & 0.1133          & 0.0943         & 0.2249          & 0.3208          & 0.1794          & 0.2995         & 0.5255         & 0.2416          \\
                      & ExpoMF                  & 0.1462          & 0.1480          & 0.1454         & 0.2900          & 0.3623          & 0.2483          & 0.3560         & 0.5569         & 0.3073          \\
                      & Rel-MF                  & 0.1470          & 0.1494          & 0.1461         & 0.3258          & 0.3870          & 0.2744          & 0.4038         & 0.5963         & 0.3482          \\
                      \cline{2-11}
                      & C-IPS           & \textbf{0.1706}          & \textbf{0.1660}         & \textbf{0.1698}         & \textbf{0.3445}          & \textbf{0.3975}          & \textbf{0.2999}          & \textbf{0.4187}         & \textbf{0.5995}         & \textbf{0.3724}          \\
                  
                      \hline\hline
\end{tabular}
\label{tab_yahoo}
\end{table*}

\section{Experiments}
\subsection{Experimental Setup}
\textbf{Datasets.}
We follow the previous work~\cite{Rel} to use \textbf{Yahoo! R3}~\cite{Yahoo} as the experiment dataset.
It is an explicit feedback dataset with a missing-not-at-random (MNAR) training set and a missing-at-random (MAR) testing set collected from a song recommendation service. In the MNAR training set, there are 311,704 five-star ratings from 15,400 users and 1,000 songs, while there are 54,000 ratings from 5,400 users on 10 randomly selected songs in the MAR testing set. Following the existing work~\cite{Rel}, we transform it into an implicit feedback dataset by treating items rated greater than or equal to 4 as positive samples, and the others as unlabeled samples. 
Basically, debiased recommender models aim to improve the performance of the user-item pairs with lower observation probabilities.
{Thus we evaluate different models based on all the testing items and the non-popular items (we remain 500 items according to their interaction frequency in the training set), respectively.}

\textbf{Baselines.}
We compare our model with the following representative baselines:
Matrix Factorization (MF)~\cite{MF} is an early well known recommender model, which has been compared in lots of previous work.
Weighted Matrix Factorization (WMF)~\cite{WMF} is a weighted matrix factorization model for tuning the importances of different samples.
Exposure Matrix Factorization (ExpoMF)~\cite{ExpoMF} is a probabilistic recommender model for simultaneously capturing the user preferences and item exposure probabilities.
Relevance Matrix Factorization (Rel-MF)~\cite{Rel} is the state-of-the-art debiased recommender model. 

\textbf{Evaluation Protocols.}
We use three metrics, including Discounted Cumulative Gain (DCG), Recall and Mean Average Precision (MAP), to evaluate the ranking performance of different models. We report the experimental results on top-$K$ recommendations by setting $K$ as $\{1, 3, 5\}$, respectively~\cite{Rel}.
Yahoo contains a MNAR set and a MAR set. We further randomly splilt the MNAR set into a training set and a validation set with the ratio $9:1$, and directly use the MAR set for testing. Since the validation set is biased, we use the self-normalized inverse propensity score estimator~\cite{SNIPS} to compute evaluation metrics for hyper-parameter tuning and model selection as existing work~\cite{RecSys_lqy18,Rel}. 

\textbf{Implementation Details.}
For the baseline algorithms, we directly use the published source code of Rel-MF\footnote{\textcolor{black}{https://github.com/usaito/unbiased-implicit-rec-real}}, which includes implementations of MF, WMF, ExpoMF and Rel-MF. We strictly follow the corresponding configurations described in the paper~\cite{Rel} to reproduce the experimental results.
For our proposed C-IPS, we implement it with TensorFlow~\cite{tensorflow2015-whitepaper} and optimize it with mini-batch Adam optimizer~\cite{kingma2014adam}. 
{$\phi$ is implemented with a two-layer fully connected neural network.}
We strictly follow Rel-MF~\cite{Rel} to configure the user/item embedding size and the learning rate.
The number of clusters $K$ is determined in the range of $\{2, 4, 6, 8, 10\}$. 

\subsection{Overall Comparison}
In this section, we evaluate our model by comparing it with the previous work. 
From the results presented in Table~\ref{tab_yahoo}, we can see:
by assigning different sample weights or modeling the item exposure probabilities, WMF and ExpoMF perform better than the basic MF model.
Rel-MF can obtain better performance than ExpoMF on both datasets, which agrees with the previous work, and demonstrates the effectiveness of the debiasing idea. 
Encouragingly, our model can consistently achieve the best performance on all the evaluation metrics across different datasets.
In specific, our model improves the best baseline by about 14.26\%, 12.94\% and 8.91\% on the metrics of DCG@1, DCG@3 and DCG@5, respectively.
These observations are not surprising, and demonstrate the effectiveness of our neural stratification idea.
We speculate that by aggregating the user interactions in the same cluster, the sample density is greatly enlarged, which can indeed help the propensity score estimation.
The improved propensity scores finally lead to more effective debiasing results, and better recommendation performance.

\subsection{Study of User Cluster Number $K$}
\begin{figure}[t]
\centering
\setlength{\fboxrule}{0.pt}
\setlength{\fboxsep}{0.pt}
\fbox{
\includegraphics[width=.5\linewidth]{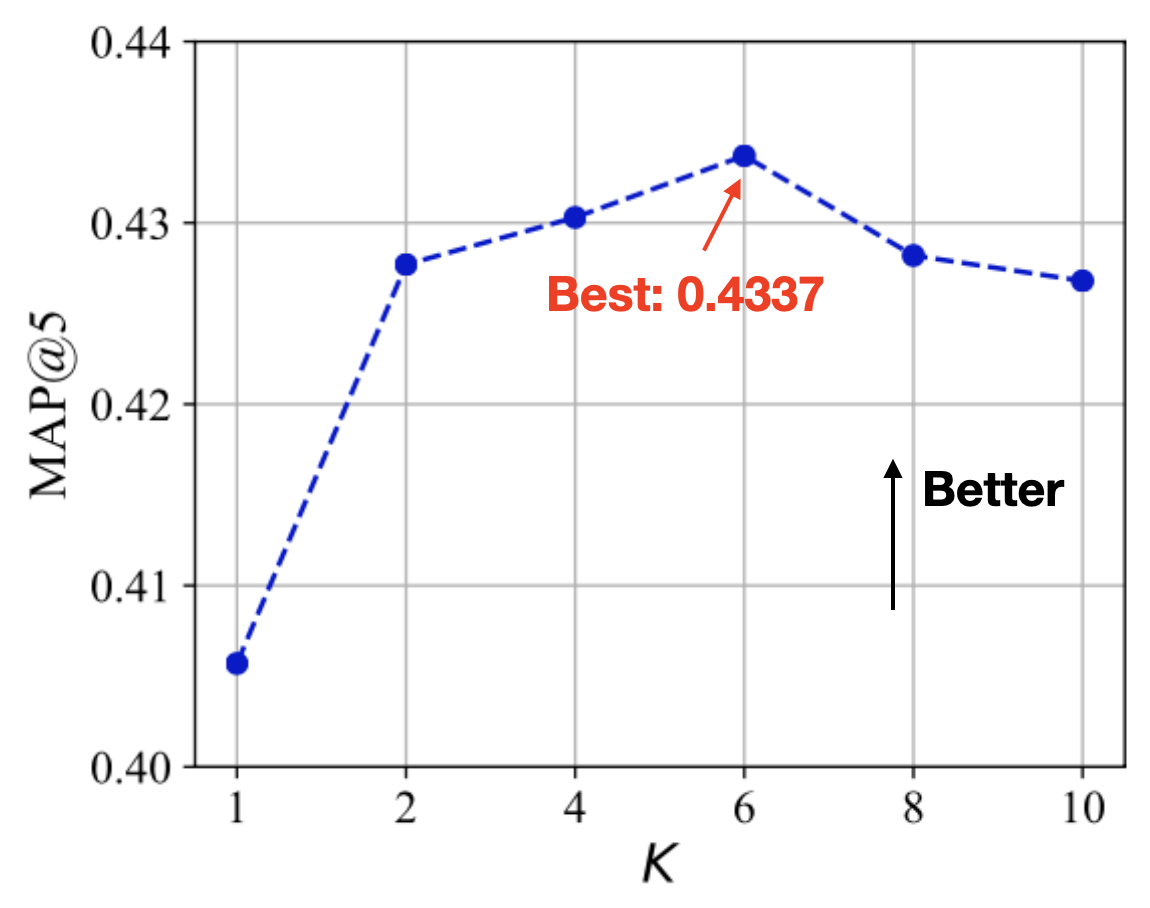}
}
\caption{Influence of $K$ on the model performance.}
\label{fig:param}
\end{figure}

An important parameter in our model is the number of clusters.
In this section, we study how the cluster number $K$ influence the model performance.
In specific, we conduct experiment based on all the testing items, and the parameters follow the above settings.
We report the results based on MAP@5, which is presented in Figure~\ref{fig:param}.
We can see the performance continually goes up as we use more clusters.
After achieving the optimal point (\emph{i.e.}, $K=6$), the performance drops significantly.
We speculate that when the cluster number is small, the user preference cannot be well decomposed and represented, which may lead to inaccurate IPS estimation.
However, if we set too much clusters, the user intrinsic preference cannot be sufficiently enhanced, and the cluster-item samples are still too sparse, which also hurts the accuracy of IPS.

\section{Related Work}
This paper aims to design a debiased recommender model.
In this field, there are a lot of previous work.
In general, the mainstream methods are mostly based on the inverse propensity scores.
For example,~\cite{IPS} proposes to build unbiased recommender model tailored for user explicit feedback.
~\cite{Rel} extends~\cite{IPS} to capture the user implicit feedback, which is more accessible and widely exists in real-world settings.
~\cite{DRJL} proposes a doubly robust method for more effective debiased recommender models.
~\cite{AutoDebias} designs a model to automatically learn the bias from the data, and accordingly debias the recommender model.
In addition to the above IPS based models, there is a recent work~\cite{CBR} on debiasing recommender model based on representation learning, where the user feature distributions are enforced to be balanced given different items.
While the above models have achieved many successes, they neither consider the latent low rank user preference structures, nor cluster users to achieve more accurate IPS estimation.

\section{Conclusion}
In this paper, we propose a novel debiased recommender model based on neural stratification.
The key idea is to cluster the users for accurately estimating the propensity scores, so as to obtain more effective debiasing models.
We empirically demonstrate the effectiveness of our model.
Actually, this paper opens the door of building debiased recommender models based on confounder clustering.
We argue that this is an interesting direction which builds a bridge between the ordinary IPS and balancing based models.
There is much room left for improvement.
To begin with, we plan to extend our framework to more comprehensive base models. 
In addition, it would also be interesting to explore different IPS computation methods based on the cluster-item exposure matrix.

\bibliographystyle{ACM-Reference-Format}
\bibliography{reference}


\begin{thebibliography}{18}


\ifx \showCODEN    \undefined \def \showCODEN     #1{\unskip}     \fi
\ifx \showDOI      \undefined \def \showDOI       #1{#1}\fi
\ifx \showISBNx    \undefined \def \showISBNx     #1{\unskip}     \fi
\ifx \showISBNxiii \undefined \def \showISBNxiii  #1{\unskip}     \fi
\ifx \showISSN     \undefined \def \showISSN      #1{\unskip}     \fi
\ifx \showLCCN     \undefined \def \showLCCN      #1{\unskip}     \fi
\ifx \shownote     \undefined \def \shownote      #1{#1}          \fi
\ifx \showarticletitle \undefined \def \showarticletitle #1{#1}   \fi
\ifx \showURL      \undefined \def \showURL       {\relax}        \fi
\providecommand\bibfield[2]{#2}
\providecommand\bibinfo[2]{#2}
\providecommand\natexlab[1]{#1}
\providecommand\showeprint[2][]{arXiv:#2}

\bibitem[\protect\citeauthoryear{Abadi, Agarwal, Barham, et~al\mbox{.}}{Abadi
  et~al\mbox{.}}{2015}]%
        {tensorflow2015-whitepaper}
\bibfield{author}{\bibinfo{person}{Mart\'{\i}n Abadi}, \bibinfo{person}{Ashish
  Agarwal}, \bibinfo{person}{Paul Barham}, {et~al\mbox{.}}}
  \bibinfo{year}{2015}\natexlab{}.
\newblock \bibinfo{title}{{TensorFlow}: Large-Scale Machine Learning on
  Heterogeneous Systems}.
\newblock
\newblock


\bibitem[\protect\citeauthoryear{Chen, Dong, Qiu, He, Xin, Chen, Lin, and
  Yang}{Chen et~al\mbox{.}}{2021}]%
        {AutoDebias}
\bibfield{author}{\bibinfo{person}{Jiawei Chen}, \bibinfo{person}{Hande Dong},
  \bibinfo{person}{Yang Qiu}, \bibinfo{person}{Xiangnan He},
  \bibinfo{person}{Xin Xin}, \bibinfo{person}{Liang Chen},
  \bibinfo{person}{Guli Lin}, {and} \bibinfo{person}{Keping Yang}.}
  \bibinfo{year}{2021}\natexlab{}.
\newblock \showarticletitle{AutoDebias: Learning to Debias for Recommendation}.
  In \bibinfo{booktitle}{\emph{SIGIR}}.
\newblock


\bibitem[\protect\citeauthoryear{Chen, Dong, Wang, Feng, Wang, and He}{Chen
  et~al\mbox{.}}{2020}]%
        {chen2020bias}
\bibfield{author}{\bibinfo{person}{Jiawei Chen}, \bibinfo{person}{Hande Dong},
  \bibinfo{person}{Xiang Wang}, \bibinfo{person}{Fuli Feng},
  \bibinfo{person}{Meng Wang}, {and} \bibinfo{person}{Xiangnan He}.}
  \bibinfo{year}{2020}\natexlab{}.
\newblock \showarticletitle{Bias and Debias in Recommender System: A Survey and
  Future Directions}.
\newblock \bibinfo{journal}{\emph{arXiv preprint arXiv:2010.03240}}
  (\bibinfo{year}{2020}).
\newblock


\bibitem[\protect\citeauthoryear{Hu, Koren, and Volinsky}{Hu
  et~al\mbox{.}}{2008}]%
        {WMF}
\bibfield{author}{\bibinfo{person}{Yifan Hu}, \bibinfo{person}{Yehuda Koren},
  {and} \bibinfo{person}{Chris Volinsky}.} \bibinfo{year}{2008}\natexlab{}.
\newblock \showarticletitle{Collaborative Filtering for Implicit Feedback
  Datasets}. In \bibinfo{booktitle}{\emph{ICDM}}. \bibinfo{publisher}{{IEEE}
  Computer Society}, \bibinfo{pages}{263--272}.
\newblock


\bibitem[\protect\citeauthoryear{Kingma and Ba}{Kingma and Ba}{2015}]%
        {kingma2014adam}
\bibfield{author}{\bibinfo{person}{Diederik~P. Kingma} {and}
  \bibinfo{person}{Jimmy Ba}.} \bibinfo{year}{2015}\natexlab{}.
\newblock \showarticletitle{Adam: {A} Method for Stochastic Optimization}. In
  \bibinfo{booktitle}{\emph{ICLR}}, \bibfield{editor}{\bibinfo{person}{Yoshua
  Bengio} {and} \bibinfo{person}{Yann LeCun}} (Eds.).
\newblock


\bibitem[\protect\citeauthoryear{Koren, Bell, Volinsky, et~al\mbox{.}}{Koren
  et~al\mbox{.}}{2009}]%
        {MF}
\bibfield{author}{\bibinfo{person}{Yehuda Koren}, \bibinfo{person}{Robert
  Bell}, \bibinfo{person}{Chris Volinsky}, {et~al\mbox{.}}}
  \bibinfo{year}{2009}\natexlab{}.
\newblock \showarticletitle{Matrix factorization techniques for recommender
  systems}.
\newblock \bibinfo{journal}{\emph{Computer}} \bibinfo{volume}{42},
  \bibinfo{number}{8} (\bibinfo{year}{2009}), \bibinfo{pages}{30--37}.
\newblock


\bibitem[\protect\citeauthoryear{Liang, Charlin, McInerney, and Blei}{Liang
  et~al\mbox{.}}{2016}]%
        {ExpoMF}
\bibfield{author}{\bibinfo{person}{Dawen Liang}, \bibinfo{person}{Laurent
  Charlin}, \bibinfo{person}{James McInerney}, {and} \bibinfo{person}{David~M.
  Blei}.} \bibinfo{year}{2016}\natexlab{}.
\newblock \showarticletitle{Modeling User Exposure in Recommendation}. In
  \bibinfo{booktitle}{\emph{WWW}},
  \bibfield{editor}{\bibinfo{person}{Jacqueline Bourdeau}, \bibinfo{person}{Jim
  Hendler}, \bibinfo{person}{Roger Nkambou}, \bibinfo{person}{Ian Horrocks},
  {and} \bibinfo{person}{Ben~Y. Zhao}} (Eds.). \bibinfo{publisher}{{ACM}},
  \bibinfo{pages}{951--961}.
\newblock


\bibitem[\protect\citeauthoryear{Marlin and Zemel}{Marlin and Zemel}{2009}]%
        {Yahoo}
\bibfield{author}{\bibinfo{person}{Benjamin~M. Marlin} {and}
  \bibinfo{person}{Richard~S. Zemel}.} \bibinfo{year}{2009}\natexlab{}.
\newblock \showarticletitle{Collaborative prediction and ranking with
  non-random missing data}. In \bibinfo{booktitle}{\emph{RecSys}},
  \bibfield{editor}{\bibinfo{person}{Lawrence~D. Bergman},
  \bibinfo{person}{Alexander Tuzhilin}, \bibinfo{person}{Robin~D. Burke},
  \bibinfo{person}{Alexander Felfernig}, {and} \bibinfo{person}{Lars
  Schmidt{-}Thieme}} (Eds.). \bibinfo{publisher}{{ACM}},
  \bibinfo{pages}{5--12}.
\newblock


\bibitem[\protect\citeauthoryear{Nigam and Ghani}{Nigam and Ghani}{2000}]%
        {Co_Training}
\bibfield{author}{\bibinfo{person}{Kamal Nigam} {and} \bibinfo{person}{Rayid
  Ghani}.} \bibinfo{year}{2000}\natexlab{}.
\newblock \showarticletitle{Analyzing the Effectiveness and Applicability of
  Co-training}. In \bibinfo{booktitle}{\emph{CIKM}}.
  \bibinfo{publisher}{{ACM}}, \bibinfo{pages}{86--93}.
\newblock


\bibitem[\protect\citeauthoryear{Saito, Yaginuma, Nishino, Sakata, and
  Nakata}{Saito et~al\mbox{.}}{2020}]%
        {Rel}
\bibfield{author}{\bibinfo{person}{Yuta Saito}, \bibinfo{person}{Suguru
  Yaginuma}, \bibinfo{person}{Yuta Nishino}, \bibinfo{person}{Hayato Sakata},
  {and} \bibinfo{person}{Kazuhide Nakata}.} \bibinfo{year}{2020}\natexlab{}.
\newblock \showarticletitle{Unbiased Recommender Learning from
  Missing-Not-At-Random Implicit Feedback}. In
  \bibinfo{booktitle}{\emph{WSDM}}.
\newblock


\bibitem[\protect\citeauthoryear{Schnabel, Swaminathan, Singh, Chandak, and
  Joachims}{Schnabel et~al\mbox{.}}{2016}]%
        {IPS}
\bibfield{author}{\bibinfo{person}{Tobias Schnabel}, \bibinfo{person}{Adith
  Swaminathan}, \bibinfo{person}{Ashudeep Singh}, \bibinfo{person}{Navin
  Chandak}, {and} \bibinfo{person}{Thorsten Joachims}.}
  \bibinfo{year}{2016}\natexlab{}.
\newblock \showarticletitle{Recommendations as Treatments: Debiasing Learning
  and Evaluation}. In \bibinfo{booktitle}{\emph{ICML}}.
  \bibinfo{pages}{1670–1679}.
\newblock


\bibitem[\protect\citeauthoryear{Swaminathan and Joachims}{Swaminathan and
  Joachims}{2015}]%
        {SNIPS}
\bibfield{author}{\bibinfo{person}{Adith Swaminathan} {and}
  \bibinfo{person}{Thorsten Joachims}.} \bibinfo{year}{2015}\natexlab{}.
\newblock \showarticletitle{The Self-Normalized Estimator for Counterfactual
  Learning}. In \bibinfo{booktitle}{\emph{NIPS}}. \bibinfo{pages}{3231--3239}.
\newblock


\bibitem[\protect\citeauthoryear{Wang, Zhang, Sun, and Qi}{Wang
  et~al\mbox{.}}{2019}]%
        {DRJL}
\bibfield{author}{\bibinfo{person}{Xiaojie Wang}, \bibinfo{person}{Rui Zhang},
  \bibinfo{person}{Yu Sun}, {and} \bibinfo{person}{Jianzhong Qi}.}
  \bibinfo{year}{2019}\natexlab{}.
\newblock \showarticletitle{Doubly Robust Joint Learning for Recommendation on
  Data Missing Not at Random}. In \bibinfo{booktitle}{\emph{ICML}}.
\newblock


\bibitem[\protect\citeauthoryear{Xie, Girshick, and Farhadi}{Xie
  et~al\mbox{.}}{2016}]%
        {DEC}
\bibfield{author}{\bibinfo{person}{Junyuan Xie}, \bibinfo{person}{Ross~B.
  Girshick}, {and} \bibinfo{person}{Ali Farhadi}.}
  \bibinfo{year}{2016}\natexlab{}.
\newblock \showarticletitle{Unsupervised Deep Embedding for Clustering
  Analysis}. In \bibinfo{booktitle}{\emph{ICML}}. \bibinfo{pages}{478--487}.
\newblock


\bibitem[\protect\citeauthoryear{Yang, Cui, Xuan, Wang, Belongie, and
  Estrin}{Yang et~al\mbox{.}}{2018}]%
        {RecSys_lqy18}
\bibfield{author}{\bibinfo{person}{Longqi Yang}, \bibinfo{person}{Yin Cui},
  \bibinfo{person}{Yuan Xuan}, \bibinfo{person}{Chenyang Wang},
  \bibinfo{person}{Serge Belongie}, {and} \bibinfo{person}{Deborah Estrin}.}
  \bibinfo{year}{2018}\natexlab{}.
\newblock \showarticletitle{Unbiased Offline Recommender Evaluation for
  Missing-Not-at-Random Implicit Feedback}. In
  \bibinfo{booktitle}{\emph{RecSys}}. \bibinfo{pages}{279–287}.
\newblock


\bibitem[\protect\citeauthoryear{Yang, Cai, Liu, Dong, He, Hao, Wang, and
  Chen}{Yang et~al\mbox{.}}{2022}]%
        {CBR}
\bibfield{author}{\bibinfo{person}{Mengyue Yang}, \bibinfo{person}{Guohao Cai},
  \bibinfo{person}{Furui Liu}, \bibinfo{person}{Zhenhua Dong},
  \bibinfo{person}{Xiuqiang He}, \bibinfo{person}{Jianye Hao},
  \bibinfo{person}{Jun Wang}, {and} \bibinfo{person}{Xu Chen}.}
  \bibinfo{year}{2022}\natexlab{}.
\newblock \showarticletitle{Debiased Recommendation with User Feature
  Balancing}.
\newblock \bibinfo{journal}{\emph{CoRR}}  \bibinfo{volume}{abs/2201.06056}
  (\bibinfo{year}{2022}).
\newblock


\bibitem[\protect\citeauthoryear{Yuan, Hsia, Yang, Zhu, Chang, Dong, and
  Lin}{Yuan et~al\mbox{.}}{2019}]%
        {CIKM_YuanHYZCDL19}
\bibfield{author}{\bibinfo{person}{Bo{-}Wen Yuan}, \bibinfo{person}{Jui{-}Yang
  Hsia}, \bibinfo{person}{Meng{-}Yuan Yang}, \bibinfo{person}{Hong Zhu},
  \bibinfo{person}{Chih{-}Yao Chang}, \bibinfo{person}{Zhenhua Dong}, {and}
  \bibinfo{person}{Chih{-}Jen Lin}.} \bibinfo{year}{2019}\natexlab{}.
\newblock \showarticletitle{Improving Ad Click Prediction by Considering
  Non-displayed Events}. In \bibinfo{booktitle}{\emph{CIKM}}.
  \bibinfo{publisher}{{ACM}}, \bibinfo{pages}{329--338}.
\newblock


\bibitem[\protect\citeauthoryear{Zhang, Bao, Liu, Yang, Lin, Wen, and
  Ramezani}{Zhang et~al\mbox{.}}{2020}]%
        {Multi_IPW}
\bibfield{author}{\bibinfo{person}{Wenhao Zhang}, \bibinfo{person}{Wentian
  Bao}, \bibinfo{person}{Xiao{-}Yang Liu}, \bibinfo{person}{Keping Yang},
  \bibinfo{person}{Quan Lin}, \bibinfo{person}{Hong Wen}, {and}
  \bibinfo{person}{Ramin Ramezani}.} \bibinfo{year}{2020}\natexlab{}.
\newblock \showarticletitle{Large-scale Causal Approaches to Debiasing
  Post-click Conversion Rate Estimation with Multi-task Learning}. In
  \bibinfo{booktitle}{\emph{WWW}}.
\newblock


\end{thebibliography}
\end{document}